\documentclass[prd,aps,showpacs,tightline,twocolumn,nofootinbib]{revtex4}
\usepackage{graphicx}
\usepackage{amssymb}
\usepackage{amsmath}

\begin{document}

\title{Nonthermal dark matter in mirage mediation}
\author{Minoru Nagai}
\author{Kazunori Nakayama}
\affiliation{Institute for Cosmic Ray Research,
University of Tokyo,
Kashiwa 277-8582, Japan}
\date{\today}

\begin{abstract}
In mirage-mediation models there exists a modulus field whose mass 
is O(1000) TeV and its late-decay may significantly change the
standard thermal relic scenario of the dark matter.
We study nonthermal production of the dark matter directly from the modulus decay,
and find that for some parameter regions non-thermally produced neutralinos
can become the dark matter.
\end{abstract}

\maketitle

\section{Introduction}

Supersymmetry (SUSY) \cite{Nilles:1983ge} 
is the best motivated physics beyond the standard model,
as it provides a solution to the hierarchy problem, 
a natural candidate of the dark matter of the universe,
and leads to the unification of the gauge coupling constants.
But SUSY must be broken spontaneously in some sector hidden from the
observable sector, and the effect of SUSY breaking must be 
communicated to the observable sector by some mediation mechanism.
There are many proposed SUSY breaking and mediation mechanisms,
such as gravity-mediation, gauge-mediation \cite{Dine:1994vc,Giudice:1998bp} 
and anomaly-mediation \cite{Randall:1998uk}.

On the other hand, supergravity in four space-time dimensions 
is believed to be low-energy effective theory of the more fundamental higher-dimensional theory, 
superstring theory \cite{Polchinski}.
One of the fundamental problems to obtain such a four dimensional theory from string theory
is that there appear many moduli which are not stabilized
when the extra dimensions are compactified.
Also the light moduli cause cosmological disaster related to the 
huge entropy production from the late-time modulus decay
\cite{Coughlan:1983ci,Banks:1993en}
or the runaway behavior of the dilaton \cite{Brustein:1992nk}.

Recently, Kachru-Kallosh-Linde-Trivedi (KKLT) \cite{Kachru:2003aw}
provided a concrete example of the 
model construction which stabilizes all the moduli. 
Motivated by these recent developments, 
low-energy consequences of KKLT-type moduli stabilization mechanism were investigated
\cite{Choi:2004sx,Endo:2005uy,Choi:2005uz,Falkowski:2005ck}.
It is found that this type of models leads to 
the mixture of modulus- and anomaly-mediation contributions to the SUSY breaking parameters,
and such models are called mirage-mediation or mixed-modulus-anomaly-mediation models.
Phenomenologically, there are some interesting features in mirage-mediation models.
One of the good points in mirage-mediation models is that they do not suffer from extra CP phases between the contributions from modulus- and anomaly-mediation.
Another is that because of the heavy gravitino mass ($m_{3/2}\sim O(10-100)$ TeV),
the cosmological gravitino problem \cite{Pagels:1981ke,Ellis:1982yb,Khlopov:1984pf}
may be significantly relaxed.
For some model parameters, the little hierarchy problem can also be relaxed
\cite{Choi:2005hd}.

However, as for the cosmological issues, there remains other problems to be solved.
In mirage-mediation models, there exists a modulus field with mass of order $\sim (8\pi^2) m_{3/2}$
and whose interactions with other particles are suppressed by the Planck scale.
It is likely that the energy stored in the coherent oscillation of the modulus field
once dominates the universe, and reheats the universe with very low reheating temperature
order $\sim O(100)$ MeV releasing huge entropy.
This invalidates the standard thermal relic dark matter scenario, 
and also significantly dilutes the preexisting baryon asymmetry. 
Thus there is a need for investigating a consistent cosmological scenario 
which can account for the present dark matter and baryon number of the universe
in the presence of such a heavy modulus field.

In order to obtain correct amount of baryon asymmetry,
a large baryon number density should be generated which can even survive the 
dilution from the modulus decay,
and perhaps the only known mechanism which can create such a large baryon number is
the Affleck-Dine mechanism \cite{Affleck:1984fy}.
In Ref.~\cite{Kawasaki:2007yy}, it is found that Affleck-Dine baryogenesis mechanism can create
the desired amount of baryon asymmetry in the presence of large dilution.

In this paper, we study the possibility of nonthermal production of the dark matter 
directly from the modulus decay and
find that the enhancement of the annihilation cross section due to the s-channel Higgs resonance
makes it possible to account for the correct relic abundance of the dark matter for some parameter regions.
A solution to the moduli-induced gravitino problem \cite{Endo:2006zj} is also discussed.
A similar subject was studied in Ref.~\cite{Moroi:1999zb} in the framework of anomaly-mediated
SUSY breaking scenario.
Our present work can be regarded as the extension of their work to the mirage-mediation model,
where the presence of moduli is necessary and inevitable.

This paper is organized as follows.
In Sec.~\ref{sec:mirage} the properties of mirage-mediation models are briefly described.
In Sec.~\ref{sec:cosmology}, the cosmology with heavy moduli is discussed.
In particular, the nonthermal dark matter from modulus decay
and its detection possibility at on-going and future direct detection experiments are investigated.
In Sec.~\ref{sec:moduliinduced} solutions to the moduli-induced gravitino problem
are proposed.
We conclude in Sec.~\ref{sec:conclusion}.

\section{A brief review on the mirage-mediation model} \label{sec:mirage}

In this section we briefly discuss the mirage-mediation models.
The K$\ddot{\rm a}$hler potential, superpotential and gauge kinetic fuction are assumed to be of the form
(hereafter we set the reduced Planck scale $M_P=1$)
\begin{equation}
\begin{split}
	&K = -3 \ln (T+T^*) + Z_i(T+T^*)\Phi_i^*\Phi_i \\
	&W = w_0-Ae^{-aT}+\frac{\lambda_{ijk}}{6}\Phi_i\Phi_j\Phi_k \\ \label{KandW}
	&f_a = kT
\end{split}
\end{equation}
where $T$ denotes the modulus field and $\Phi_i$ denotes the MSSM superfields
and $Z_i(T+T^*)=1/(T+T^*)^{n_i}$.
The subscript $i$ distinguishes the matter species, and  
$a$ and $k$ are real constants.
The value of $n_i$ are determined by the concrete model setup,
but here we regard them as free parameters which characterize the phenomenological
properties.
This type of the K$\ddot{\rm a}$hler potential and superpotential of the modulus arises from string theoretical 
construction such as the KKLT model.
The scalar potential for the modulus is given by
\begin{equation}
	V_s=e^{K(T+T^*)} \left [K^{T\bar T} |D_T W|^2-3|W|^2 \right ]
\end{equation}
where $D_T W=W_T +K_T W $ and $K^{T\bar T}=(K_{T\bar T})^{-1}$
(here the subscript $T$ denotes partial derivative with it).
The resulting minimum of this potential is a supersymmetric AdS vacuum.
In order to obtain dS vacuum, we add the additional uplifting potential 
from the brane sequestered from the observable sector, 
\begin{equation}
	V_{\rm lift}=\frac{D}{(T+T^*)^m}
\end{equation}
where $m$ is $O(1)$ constant.
Then the total scalar potential is given by $V=V_s+V_{\rm lift}$.
After fine-tuning the value of $D$, we can obtain correct dS vacuum with 
positive vacuum energy density observed now.

This model has interesting phenomenological consequences.
There are two sources of SUSY breaking in the present model.
One comes from the uplifting brane, which gives the dominant contribution to the SUSY breaking
and determines the mass of gravitino.
On the other hand, at the dS minimum after uplifting the modulus has 
non-vanishing $F$-term. 
This also gives SUSY breaking effect.
Roughly speaking, the resulting SUSY breaking parameters are determined by the combination of
anomaly-mediated SUSY breaking effect which comes from uplifting brane
and modulus-mediated SUSY breaking effect which comes from the modulus itself.

The contribution to the gaugino masses, sfermion masses and $A$ terms 
from the modulus-mediation at GUT (Grand Unified Theory) scale are denoted as
 $M_0, \tilde m_i^2$ and $\tilde A_{i}$,
 which are related to $n_i$.
Then we define the following parameters which distinguish mirage-mediation models,
$a_i = \tilde A_i/M_0$, $c_i=\tilde m_i^2/M_0^2$ and 
\begin{equation}
	\alpha = \frac{m_{3/2}}{M_0 \ln (M_P/m_{3/2})}
\end{equation}
which describes the ratio of the anomaly-mediation to modulus-mediation contribution
to the SUSY breaking parameters.
Thus mirage-mediation models are parameterized by the following parameters
\cite{Choi:2005uz},
\begin{equation}
 	M_0,c_i,a_i,\tan \beta, \alpha 
\end{equation}
where $\tan \beta$ represents the ratio
of the vacuum expectation values of the up-type and down-type Higgs,
$ \langle H_u \rangle /\langle H_d \rangle$.
We assume $\mu > 0$, as positive $\mu$ is favored from the muon $g-2$ experiment.

When the renormalization group equations are applied to obtain low-energy mass spectrum,
one finds that the gaugino masses at the energy scale $\mu$ is given by
\begin{equation}
	M_a(\mu) = \frac{g_a^2(\mu)}{g_a^2(M_{\rm mir})}M_0,
\end{equation}
where 
\begin{equation}
	M_{\rm mir}=\frac{M_{\rm GUT}}{(M_P/m_{3/2})^{\alpha/2}}
\end{equation}
with $M_{\rm GUT} \simeq 2\times 10^{16}~$GeV.
Thus they are unified at the scale $\mu=M_{\rm mir}$,
which is called the mirage-scale,
although there are no new physics at this scale.
Note that gauge coupling constants unification at the GUT scale is not modified.
For $\alpha =1$, the mirage-scale actually becomes intermediate scale,
$M_{\rm mir} \sim 10^9$ GeV.
For $\alpha =2$, it becomes $M_{\rm mir} \sim 1$ TeV.
This case is called TeV scale mirage-mediation.

One of the interesting properties of mirage-mediation models is that
there exists a modulus field whose interaction is Planck-suppressed.
It is found that the modulus mass around its potential minimum is
estimated as $m_T \sim (8\pi^2)m_{3/2}$.
Because the gravitino mass is one or two-orders of magnitude larger than the 
other SUSY particle masses, the modulus mass is expected to be $m_T \gg O(100)$ TeV.
The modulus with such a large mass decays well before Big-Bang nucleosynthesis (BBN)
and hence it seems harmless for cosmology.
But the typical decay temperature of the modulus is much lower than the
freeze-out temperature of the lightest supersymmetric particles (LSPs),
which invalidates the standard thermal relic scenario.
In the next section, we consider the nonthermal production of the LSPs directly from the
modulus decay
\footnote{Another problem is huge dilution of the preexisting baryon asymmetry
due to the modulus decay. 
Affleck-Dine baryogenesis \cite{Affleck:1984fy,Fujii:2001zr,Kawasaki:2006yb}
can surely create enough baryon number 
which survives the dilution \cite{Dolgov:2002vf,Kawasaki:2007yy}. }.

\section{Nonthermal neutralino dark matter} \label{sec:cosmology}

After inflation ends, the modulus is likely to dominate the universe before the decay
because of its large energy density stored in the form of scalar condensates
with large initial amplitude of order the Planck scale.
Thus the final radiation dominated regime is realized after the modulus decay.
The final reheating temperature is estimated as
\begin{equation}
	T_T \sim 170~{\rm MeV} \sqrt c \left ( \frac{m_T}{10^3~{\rm TeV}} \right )^{3/2}
\end{equation}
where we have used the decay rate of the moduli,
$\Gamma_T = cm_T^3/4\pi M_P^2$ with $O(1)$ constant $c$.
This satisfies the lower limit on the reheating temperature from BBN
\cite{Kawasaki:1999na,Hannestad:2004px,Ichikawa:2005vw},
but is well below the typical freeze-out temperature of the LSP, 
$T_f \sim m_{\rm L}/20 \gtrsim O(10)$ GeV
where $m_{\rm L}$ denotes the mass of the LSP.
Thus we can neglect the production process of the LSPs 
in the thermal bath around the epoch of the modulus decay.
The relevant Boltzmann equations for the LSP, modulus and radiation are
\begin{gather}
	\dot n_{\rm L}+3Hn_{\rm L} = -\langle \sigma v \rangle n_{\rm L}^2
	+2B_{\tilde \chi}\Gamma_T n_T ,\\
	\dot n_{T}+3Hn_{T}= -\Gamma_T n_T,\\
	\dot \rho_{\rm rad}+4H\rho_{\rm rad}=(m_T-2B_{\tilde \chi}m_{\rm L})\Gamma_T n_T
	+m_{\rm L} \langle \sigma v \rangle n_{\rm L}^2
\end{gather}
where $\langle \sigma v \rangle $ denotes the thermally averaged annihilation cross section
of the LSP
\footnote{
Neutralinos are expected to soon reach kinetic equilibrium for $T_T\gtrsim O(10)$ MeV due to scattering with background particles in thermal bath 
\cite{Kawasaki:1995cy,Hisano:2000dz}.
},
$n_{\rm L}$ and $n_T$ denote the number density of the LSP and modulus,
$\rho_{\rm rad}$ denotes the radiation energy density,
$H$ is the Hubble parameter,
and $B_{\tilde \chi}$ denotes the branching ratio of the modulus decay into gauginos,
which is roughly the same order as that into gauge bosons.
For $t >\tau_T$ where $\tau_T$ is the lifetime of the moduli,
the evolution of the LSP number density is well described by
\begin{equation}
	 \dot n_{\rm L}+3Hn_{\rm L} = -\langle \sigma v \rangle n_{\rm L}^2.
\end{equation}
Changing the variable to $Y=n_{\rm L}/s$ where $s$ is the entropy density,
this equation can be integrated under the assumption that the relativistic effective degrees of
freedom ($g_*$) remains constant.
As a result, we obtain \cite{Fujii:2001xp}
\begin{equation}
	Y(T) = \left[ \frac{1}{Y(T_T)}+\sqrt{\frac{8\pi^2 g_*}{45}}
	\langle \sigma v \rangle M_P(T_T-T)  \right]^{-1}.   \label{Ynonthermal}
\end{equation}
The initial abundance of the LSP $Y(T_T)$ is given as
\begin{equation}
	Y(T_T)=2B_{\tilde \chi}\frac{3T_T}{4m_T}.
\end{equation}
Thus in the relevant parameter region, the second term in Eq.~(\ref{Ynonthermal})
always dominates and the final relic abundance of the LSP is 
inversely proportional to the annihilation cross section similar to the usual thermal relic scenario
\cite{Moroi:1994rs,Kawasaki:1995cy,Moroi:1999zb,Fujii:2001xp,Gelmini:2006pw}.
The difference is that the larger annihilation cross section is required 
to obtain the desired amount of dark matter since the decay temperature of the modulus
is much lower than the freeze-out temperature of the LSP.

Then, whether the desired amount of dark matter is obtained or not 
crucially depends on the properties of the LSP.
In mirage-mediation models, the lightest neutralino naturally becomes the LSP
and it is the mixture of bino and higgsino.
For small $\alpha$, where the modulus-mediation effect dominates, 
the lightest neutralino is mainly bino-like.
As the value of $\alpha$ is increased, the anomaly-mediation contribution
reduces the mass of the gluino, which leads to lighter stop mass and smaller $|\mu|$ value.
Thus for large $\alpha$ the higgsino-like LSP is also viable.

\begin{figure}[htbp]
	\begin{center}
		\includegraphics[width=1.0\linewidth]{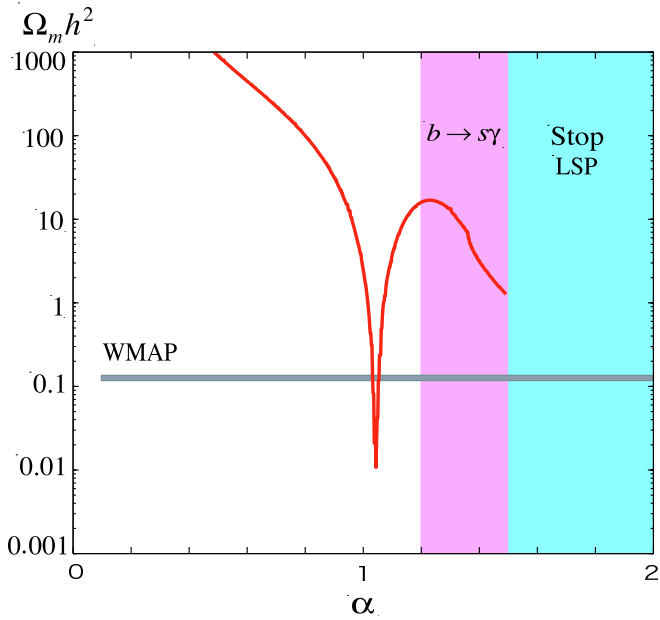}
		\caption{The abundance of non-thermally produced neutralinos $\Omega_mh^2$
		as a function of $\alpha$,
		for $M_0=350$ GeV, $c_M=a_M=1$, $c_H=a_H=0$ and $\tan \beta = 10$. }
		\label{fig:oh2-m350}
	\end{center}
\end{figure}

The result is shown in Fig.~\ref{fig:oh2-m350}
for $M_0=350$ GeV, $c_M=a_M=1$, $c_H=a_H=0$ and $\tan \beta = 10$,
where the subscript $H$ denotes the Higgs fields and 
$M$ denotes other matter fields.
In the following analysis, we have used modified DarkSUSY routine \cite{Gondolo:2004sc}
to calculate the
annihilation cross section and spin-independent neutralino-nucleon scattering cross section.
The purple shaded region is excluded from the $b\to s \gamma$ constraint.
In the blue shaded region stop becomes the LSP.
The gray band shows the favored region from the observation of 
WMAP three year results \cite{Spergel:2006hy}.
In the calculation of the relic abundance of the dark matter, 
only the S-wave contributions are included since the velocity of dark matters is very tiny and its effect on the annihilation cross section can be safely neglected.
The coannihilation effects are also less important 
since the coannihilating particles decay very quickly compared to the rate of annihilation processes.
Thus they are not included in our analysis.

In general, the bino has small annihilation cross section and is not suitable candidate
of the nonthermal dark matter.
But in mirage-mediation models there are relatively large parameter regions
where an enhancement from the s-channel Higgs resonance (also called $A$-funnel)
is obtained.
Due to this enhancement, the bino-like neutralino can be the interesting candidate
of the dark matter as can be seen from Fig.~\ref{fig:oh2-m350}.


\begin{figure}[htbp]
	\begin{center}
		\includegraphics[width=1.0\linewidth]{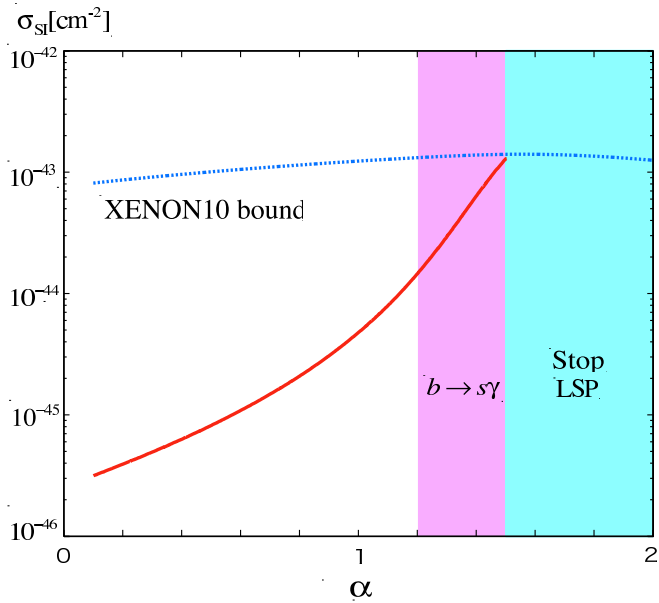}
		\caption{The spin-independent cross section of neutralinos with nucleons
		$\sigma_{\rm SI}$
		as a function of $\alpha$.
		 Parameters are the same as Fig.~\ref{fig:oh2-m350}. }
		\label{fig:SI-m350}
	\end{center}
\end{figure}

In Fig.~\ref{fig:SI-m350}, the spin-independent scattering cross section of neutralinos
with nucleons is shown for the same parameter as Fig.~\ref{fig:oh2-m350}.
For larger $\alpha$ the higgsino component in the lightest neutralino state increases
as described above, which leads to the enhancement on the 
nucleon-neutralino scattering cross section.
The recently reported bound from XENON10 experiment  \cite{Angle:2007uj} is shown 
by the dotted line.
The interesting region which accounts for the dark matter density of the universe 
satisfies the bound,
and may be within the reach of the future direct detection experiments.


\begin{figure}[htbp]
	\begin{center}
		\includegraphics[width=1.0\linewidth]{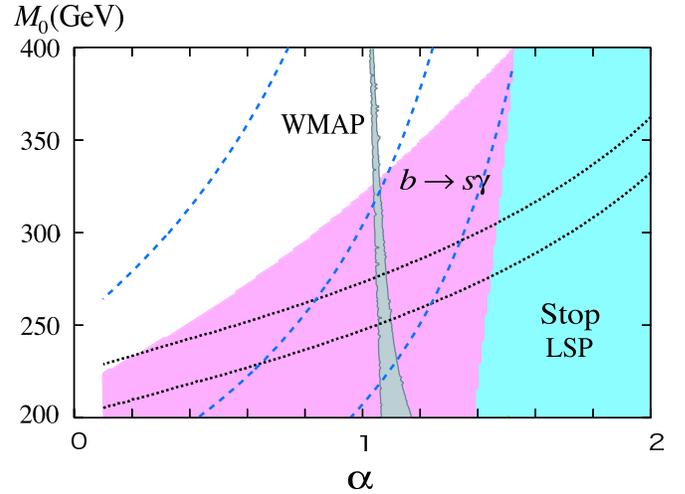}
		\caption{ 
		The  WMAP dark matter region is indicated by gray shaded region.
		The light gray shaded region predicts $\Omega_m h^2 < 0.1$.
		The upper dotted line represents $m_h=115$ GeV while
		the lower one corresponds to $m_h=114$ GeV.
		Blue dashed lines represent $\sigma_{\rm SI}=10^{-45},10^{-44},10^{-43}
		~{\rm cm}^{-2}$ from upper to lower.
		 Other parameters are the same as Fig.~\ref{fig:oh2-m350}. }
		\label{fig:alpha-mhf-oh2}
	\end{center}
\end{figure}

More large parameter regions are investigated in Fig.~\ref{fig:alpha-mhf-oh2}.
The  WMAP dark matter region is indicated by gray shaded region.
The light gray shaded region predicts too low relic abundance
to account for the dark matter of the universe, $\Omega_m h^2 < 0.1$.
The black upper dotted line represents $m_h=115$ GeV while
the lower one corresponds to $m_h=114$ GeV,
where $m_h$ denotes the lightest Higgs boson mass.
Also the neutralino-nucleon spin-independent scattering cross section is shown
by blue dashed lines, $\sigma_{\rm SI}=10^{-45},10^{-44},10^{-43}
~{\rm cm}^{-2}$ from upper to lower.
It is found that since the nonthermal scenario requires relatively light SUSY masses,
the constraints from $b\to s \gamma$ and Higgs mass are rather severe.
In this parameter regions, the SUSY contribution for $b\to s \gamma$ is dominated by the chargino one and it has the opposite sign to the standard model contribution.
Since the recent theoretical prediction of the standard model \cite{Becher:2006pu} is deviated about 1.4 $\sigma$ below the experimental value \cite{Barberio:2007cr},
 this chargino contribution severely constrains the parameters of the model.


\begin{figure}[htbp]
	\begin{center}
		\includegraphics[width=1.0\linewidth]{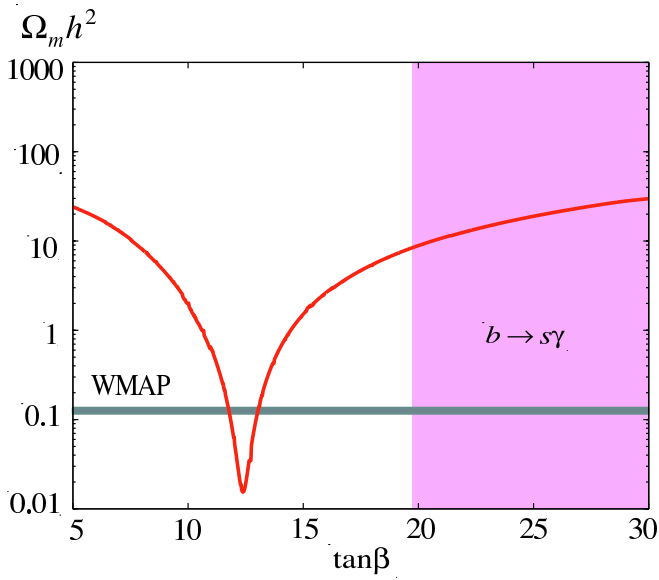}
		\caption{ 
		The abundance of non-thermally produced neutralinos $\Omega_mh^2$
		as a function of $\tan \beta$,
		for $M_0=450$ GeV, $c_M=a_M=1$, $c_H=a_H=0$ with $\alpha=1$ is fixed. }
		\label{fig:tanbe-oh2}
	\end{center}
\end{figure}

\begin{figure}[htbp]
	\begin{center}
		\includegraphics[width=1.0\linewidth]{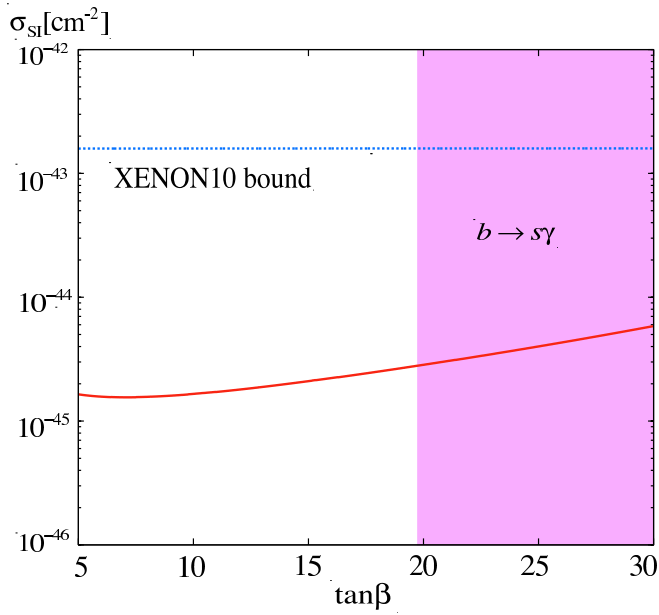}
		\caption{ 
		The spin-independent cross section of neutralinos with nucleons
		$\sigma_{\rm SI}$
		as a function of $\tan \beta$.
		 Parameters are the same as Fig.~\ref{fig:tanbe-oh2}. }
		\label{fig:tanbe-SI}
	\end{center}
\end{figure}

Also in Figs.~\ref{fig:tanbe-oh2} and \ref{fig:tanbe-SI} the results with 
$M_0=450$ GeV, $c_M=a_M=1$, $c_H=a_H=0$ and $\alpha = 1$
with varying $\tan \beta$ are shown.
Also it is found that due to the s-channel Higgs resonance, 
the resulting abundance of the non-thermally produced neutralinos can lie in the
favored range.


\begin{figure}[htbp]
	\begin{center}
		\includegraphics[width=1.0\linewidth]{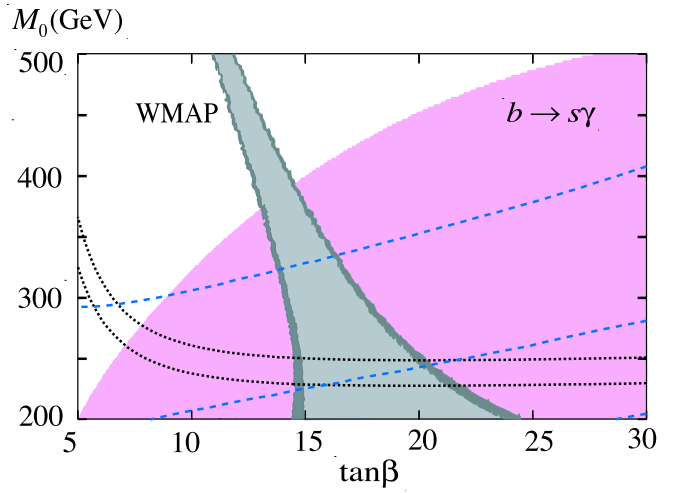}
		\caption{ 
		The  WMAP dark matter region is indicated by gray shaded region.
		The light gray shaded region predicts $\Omega_m h^2 < 0.1$.
		The upper dotted line represents $m_h=115$ GeV while
		the lower one corresponds to $m_h=114$ GeV.
		Blue dashed lines represent $\sigma_{\rm SI}=10^{-44},10^{-43},10^{-42}
		~{\rm cm}^{-2}$ from upper to lower.
		Other parameters are the same as Fig.~\ref{fig:tanbe-oh2}. }
		\label{fig:tanbe-mhf}
	\end{center}
\end{figure}

The contour plot is shown in Fig.~\ref{fig:tanbe-mhf}
in the ($\tan \beta, M_0$) plane.
In this figure $\alpha$ is fixed to 1.
Other parameters are the same as those in Fig.~\ref{fig:tanbe-oh2}.
The WMAP dark matter region is indicated by gray shaded region.
The light gray shaded region predicts too low relic abundance
to account for the dark matter of the universe.
The meanings of black dotted lines are the same as that of Fig.~\ref{fig:alpha-mhf-oh2}.
Also the neutralino-nucleon spin-independent scattering cross section is presented
by blue dashed lines, $\sigma_{\rm SI}=10^{-44},10^{-43},10^{-42}
~{\rm cm}^{-2}$ from upper to lower.
Although $b\to s \gamma$ constraint is severe for small $M_0$ and large $\tan \beta$,
the s-channel resonance effect enhances the
annihilation cross section which leads to smaller abundance of the nonthermal LSPs
in rather wide parameter range.


\section{Problems with non-thermally produced gravitinos} \label{sec:moduliinduced}

So far, we have ignored the possibly important process, the modulus decay into gravitinos.
Recently it is pointed out that the branching ratio of the modulus decay into gravitinos $B_{3/2}$
is not suppressed and in general $B_{3/2}\sim O(0.01-0.1)$
\cite{Endo:2006zj,Asaka:2006bv}.
The subsequent decay of non-thermally produced gravitinos cause another cosmological difficulty,
the destruction of light elements synthesized through BBN
and overproduction of LSPs from the gravitino decay.
Although the decay rate of moduli into gravitinos depends on the details of the SUSY breaking sector
\cite{Dine:2006ii,Endo:2006tf,Lebedev:2006qq,Lebedev:2006qc},
it is worth mentioning here the case where such a decay mode is not suppressed.

The abundance of gravitinos produced directly from the modulus decay is given by
\begin{equation}
\begin{split}
	Y_{3/2} &= 2B_{3/2} \frac{3 T_T}{4m_T} \\
	& \sim 2.6\times 10^{-9}c^{1/2}\left ( \frac{B_{3/2}}{0.01} \right )
	\left ( \frac{10.75}{g_*(T_T)} \right )^{1/2}
	\left ( \frac{m_T}{10^3~{\rm TeV}} \right )^{1/2}.
\end{split}
\end{equation}
If the gravitinos decay well before BBN, the produced LSPs overclose the universe.
If the gravitinos decay after BBN starts, the decay process disturbs BBN \cite{Kawasaki:1994af}.
Both constraints require the additional entropy production, 
which dilutes the gravitinos by an amount of $\Delta \gtrsim 10^3$
after the production of gravitinos by the modulus decay.
A concrete example of such an entropy production is
the late decay of Q-balls \cite{Kawasaki:2007yy,Fujii:2002aj}, 
which are non-topological solitons formed through
Affleck-Dine mechanism 
\cite{Coleman:1985ki,Kusenko:1997si,Enqvist:1997si,Kasuya:1999wu}.
This model requires no additional fields other than MSSM fields, and
it is interesting since the Q-ball decay not only solves the moduli-induced gravitino problem,
but also can provide the correct amount of baryon asymmetry \cite{Kawasaki:2007yy}.

Another possibility is thermal inflation caused by some additional singlet field called flaton
\cite{Yamamoto:1985rd,Lyth:1995hj}.
Thermal inflation can sufficiently dilute the modulus abundance \cite{Asaka:1999xd}
in order that
the subsequent modulus decay has no significant effects on cosmology.
Finally the flaton reheats the universe with reheating temperature of $O$(GeV) typically
\footnote{It is possible that the reheating temperature from the flaton decay is
higher than the freeze-out temperature of the LSP.
In such a case, the standard thermal relic scenario holds \cite{Baer:2006id}. }.
However, it is not obvious how the sizable amount of baryon asymmetry is generated 
in the presence of thermal inflation 
\footnote{See e.g., Refs.~\cite{Stewart:1996ai,Kasuya:2001tp} for related works.}.
 
After the dilution, non-thermally produced LSPs from the modulus decay
may have neglecting abundance.
However, in both cases, the late decay of Q-balls (or flaton) at low temperature 
also produces nonthermal LSPs \cite{Fujii:2001xp,Fujii:2003iq}.
The resulting abundance of nonthermal LSPs is given by similar manner to the case of modulus decay.
In both cases, the Q-ball (or flaton) decay into gravitinos is obviously suppressed
since their interaction is not gravitationally suppressed,
or such a decay mode can be kinematically forbidden since the gravitino is heavy enough.

\begin{figure}[htbp]
	\begin{center}
		\includegraphics[width=1.0\linewidth]{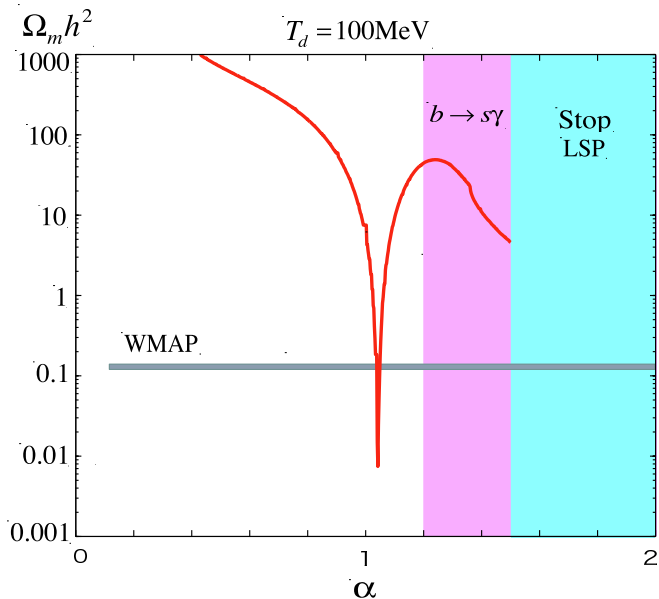}
		\caption{The abundance of non-thermally produced neutralinos 
		from the Q-ball decay $\Omega_mh^2$
		as a function of $\alpha$.
		 Parameters are the same as Fig.~\ref{fig:oh2-m350}. }
		\label{fig:oh2-m350-T100}
	\end{center}
\end{figure}

In Fig.~\ref{fig:oh2-m350-T100} the abundance of nonthermal LSPs
produced by the Q-ball (flaton) decay for the decay temperature $T_d=100$~MeV is shown.
The result is similar to that from the modulus decay.
Note that the resulting LSP abundance is inversely proportional to $T_d$,
so $T_d$ can not be as low as 10 MeV.
Contrary to the case with moduli, the decay temperature of the Q-ball (flaton) $T_d$
is not directly related to model parameters.
But in order to dilute the gravitinos produced by the modulus decay,
$T_d$ should not be so large.


\section{Conclusions } \label{sec:conclusion}

In this paper, we studied the possibility of nonthermal dark matter from the modulus decay
in mirage-mediation model,
motivated by the cosmological consideration that the modulus likely to dominate the universe
before the decay and its decay temperature is very low.
It is found that through the s-channel Higgs resonance, 
the annihilation cross section of binolike neutralino can be enhanced and 
the correct amount of dark matter can be remained.
Although the parameter region which gives such a feature is constrained
from the $b\to s \gamma$ bound and Higgs mass bound, 
the nonthermal production scenario of the dark matter from the modulus decay is viable 
for natural parameters around $\alpha \sim 1$, $\tan \beta \sim 10$
and $M_0 \sim 300-500$ GeV.

On the other hand, if the moduli decay into gravitinos with unsuppressed branching ratio, 
such gravitinos cause cosmological problems.
In order to avoid the difficulty, the additional late-time entropy production is needed.
Q-balls or thermal inflation flatons may be the possible candidates which give
sufficient entropy production.
We have also shown that 
although their late decay may invalidate the nonthermal dark matter scenario from the moduli,
their decay itself can again produce the sizable amount of dark matter non-thermally.
In the Q-ball scenario, not only the dark matter abundance
but also the baryon asymmetry of the universe can be explained simultaneously.

Our final comment is that such non-thermally produced neutralinos
can not be the warm dark matter, since
they can not freely stream without losing their energy due to the scattering with background particles.
In this respect, the missing satellite problem \cite{Klypin:1999uc}
or the cusp problem \cite{Moore:1994yx} should be solved
without referring to the nature of the dark matter (see e.g., Ref.~\cite{Bullock:2000wn}).

If any, nonthermal production scenario of the dark matter may have distinct implications
on the collider and direct/indirect dark matter detection experiments and
is interesting as a probe of the cosmological evolution scenario before BBN.
Generally, in supergravity or superstring-inspired theory,
there appear long-lived scalar fields whose late decay disturbs the standard 
thermal history of the universe.
Thus we believe it is important to seek a consistent cosmological scenario taking into account
the dynamics of such harmful scalar fields in the hidden sector.

\begin{acknowledgements}

We are grateful to J.~Hisano for useful comments.
M.N. and K.N. would like to thank the Japan Society for the Promotion of Science for financial support.

\end{acknowledgements}



\end{document}